\documentclass[aps,nopacs,nofootinbib,superscriptaddress,groupedaddress,]{revtex4}  
\usepackage[english]{babel}
\usepackage{amsmath,amssymb}
\usepackage{graphicx}
\usepackage[sort&compress]{natbib}
\usepackage{slashed}            
\allowdisplaybreaks

\bibpunct{[}{]}{,}{n}{}{,}

\def\beq{\begin{equation}}
\def\eeq{\end{equation}}
\def\beqn{\begin{eqnarray}}
\def\eeqn{\end{eqnarray}}

\def\I33m{\mathrm{I}_3^{3{\mathrm m}}}

\def\be{\begin{equation}}
\def\ee{\end{equation}}
\def\bea{\begin{eqnarray}}
\def\eea{\end{eqnarray}}

\def\spa#1.#2{\left\langle#1#2\right\rangle}
\def\spb#1.#2{\left[#1#2\right]}
\def\lor#1.#2{\left(#1#2\right)}
\def\sand#1.#2.#3{%
\left\langle\smash{#1}{\vphantom1}^{-}\right|{#2}%
\left|\smash{#3}{\vphantom1}^{-}\right\rangle}
\def\sandp#1.#2.#3{%
\left\langle\smash{#1}{\vphantom1}^{-}\right|{#2}%
\left|\smash{#3}{\vphantom1}^{+}\right\rangle}
\def\sandpp#1.#2.#3{%
\left\langle\smash{#1}{\vphantom1}^{+}\right|{#2}%
\left|\smash{#3}{\vphantom1}^{+}\right\rangle}
\def\sandpm#1.#2.#3{%
\left\langle\smash{#1}{\vphantom1}^{+}\right|{#2}%
\left|\smash{#3}{\vphantom1}^{-}\right\rangle}
\def\sandmp#1.#2.#3{%
\left\langle\smash{#1}{\vphantom1}^{-}\right|{#2}%
\left|\smash{#3}{\vphantom1}^{+}\right\rangle}
\def\spab#1.#2.#3{\langle#1|#2|#3]}
\def\spba#1.#2.#3{[#1|#2|#3\rangle}
\def\spaa#1.#2.#3{\langle#1|#2|#3\rangle}
\def\spbb#1.#2.#3{[#1|#2|#3]}
\def\spaxa#1.#2.#3.#4{\langle#1|#2|#3|#4\rangle}
\def\spbxb#1.#2.#3.#4{[#1|#2|#3|#4]}

\newcommand\Spaa[1]{\langle #1 \rangle}
\newcommand\Spab[1]{\langle #1 ]}

\newcommand*\diff{\mathop{}\!\mathrm{d}}
\newcommand*\Diff[1]{\mathop{}\!\mathrm{d^#1}}

\begin{document}

\title{$\gamma\gamma\gamma\gamma$ production at the LHC: an application of $2\rightarrow 4$ analytic unitarity}

\author{Tristan Dennen}

\affiliation{Niels Bohr International Academy and Discovery Center \\
The Niels Bohr Institute, University of Copenhagen \\
Blegdamsvej 17, DK-2100 Copenhagen \O, Denmark}
\email{tdennen@nbi.ku.dk}

\author{Ciaran Williams}
\affiliation{Niels Bohr International Academy and Discovery Center \\
The Niels Bohr Institute, University of Copenhagen \\
Blegdamsvej 17, DK-2100 Copenhagen \O, Denmark}
\affiliation{Department of Physics, University at Buffalo \\
The State University of New York,
Buffalo, NY 14260-1500, USA}
\email{ciaran@nbi.dk, ciaranwi@buffalo.edu}

\date{\today}

\begin{abstract}
We present Next-to-Leading Order (NLO) predictions for $\gamma\gamma\gamma\gamma$ final states, including the effects of photon fragmentation.
Our results are calculated fully analytically using the techniques of $D$-dimensional unitarity, and we discuss some refinements 
to existing methods, focusing particularly on the role of three-mass triangle coefficients. 
Using these techniques we are able to produce a numerically stable Monto Carlo code that runs fully in double precision.
We investigate the $4\gamma$ signal at colliders, both for Run II of the LHC and at future machines. 
Our results are implemented in MCFM. 
\end{abstract}

\maketitle

\section{Introduction} 

The production of multiple vector boson final states in hadronic collisions provides an 
opportunity to study the predictions of the Standard Model (SM) of particle physics in an interesting 
regime both theoretically and experimentally. On the theoretical side, the production relies upon 
the properties of both the underlying Electroweak (EW) gauge group and the strong interactions of QCD. Indeed 
the EW bosons, singlets under $SU(N_c)$, provide hard probes of the underlying scattering. Secondly, 
many models of new physics assume deviations from the SM predictions in the EW sector. As such, the production 
of multi-boson final states is a natural place to look for potential new physics effects. 
Experimentally the signatures associated with EW gauge bosons and their (leptonic) decay products often represent 
the cleanest and best resolved channels and therefore allow for a wide variety of interesting searches for rare processes. 

Of the EW gauge bosons, the massless photon has been studied in the most detail. There is a rich history of measurements 
of photonic processes at hadron colliders, with studies being performed at every major machine over the last few decades~\cite{Abachi:1996qz,Bonvin:1988yu,Albajar:1988im,Alitti:1992hn,Abe:1992cy,Abazov:2010ah,Aaltonen:2012jd,Chatrchyan:2011qt,Aad:2012tba,Aad:2013zba,Aaltonen:2011vk,Chatrchyan:2013mwa}.
These analyses have historically studied the high rate inclusive photon $pp\rightarrow \gamma +X$ and diphoton $pp\rightarrow \gamma\gamma +X$ processes, for which dedicated Next-to-Leading Order (NLO) Monte Carlo codes have existed for some time~\cite{Catani:2002ny,Binoth:1999qq,Bern:2001df,Bern:2002jx}. 
Recent calculations have extended the theoretical accuracy to Next-to-Next-to Leading Order~\cite{Catani:2011qz} (NNLO), improving the overall agreement 
between the experimental data and theoretical prediction in both rate and differential distributions.
 In addition to the inclusive production rates, cross sections for photon pairs in association with jets can also be studied,
for which NLO calculations for $\gamma\gamma+$jet have been completed~\cite{DelDuca:2003uz,Gehrmann:2013aga}. A comprehensive understanding of these processes
is crucial in light of the discovery of a light Higgs boson that decays to photon pairs \cite{Chatrchyan:2012ufa,Aad:2012tfa}. Of particular interest is the study of differential Higgs distributions 
across different jet bins~\cite{Aad:2014lwa}. In the effective field theory, NNLO predictions have recently become available for Higgs plus one jet~\cite{Boughezal:2013uia,Chen:2014gva}, and 
NLO predictions matched to parton showers are available for Higgs plus one or two jets~\cite{Campbell:2012am,Hoeche:2012yf,Frederix:2012ps}.  
Recently, the theoretical predictions 
for the backgrounds for these processes have been further improved by the NLO calculations of $\gamma\gamma +2j$~\cite{Gehrmann:2013bga,Bern:2014vza,Bern:2013bha,Badger:2013ava}.

Both theoretical predictions and experimental measurements are complicated by the need to define photon 
isolation requirements. Experimentally, large backgrounds associated with photons that arise from secondary decays such as $\pi^0\rightarrow \gamma\gamma$
must be suppressed. Therefore
the hadronic energy near the photon is required to be
less than either a fixed limit or a fraction of the photon transverse momentum ($p_T$).
Theoretically, at NLO the presence of collinear singularities between final state fermions and photons 
mandates the inclusion of fragmentation functions~\cite{Bourhis:1997yu,GehrmannDeRidder:1997gf}, 
which ensure the cross section is well-defined and free of infrared (IR) singularities~\cite{Catani:2002ny}. The theoretical 
need for fragmentation functions can be avoided by instead requiring that the photons satisfy an alternate form of isolation proposed 
in Ref.~\cite{Frixione:1998jh}. Using this prescription the hadronic energy is limited by a smoothly varying function, which allows arbitrarily 
soft radiation inside the bulk of the cone (needed for IR safety), but excludes the precisely collinear point, removing the fragmentation function. 
Experimentally, this type of isolation is challenging to implement, primarily due to the discrete nature of the detector calorimeter cells.

Studies of triboson production represent the cutting edge of multiple boson studies, with analyses of Run I LHC data expected to
measure several triboson processes in the near future. Of these the triphoton process represents
one of the most accessible~\cite{Bozzi:2011en,Campbell:2014yka}. Ref.~\cite{Campbell:2014yka} performed a detailed study of this process in which 
different isolation schemes were compared and contrasted. It was shown that, in general, there is reasonable agreement between the two different isolation 
prescriptions, in particular when the fixed order $\mathcal{O}(\alpha)$ fragmentation set of Ref.~\cite{GehrmannDeRidder:1997gf} was used in conjunction with the NLO prediction. By comparing 
$\gamma\gamma\gamma$ and $\gamma\gamma$+jet, the dependence on the number of photons in the final state was also studied. It was shown that the shape of the isolation 
dependence was affected most strongly by the underlying kinematics (i.e. both $2\rightarrow 3$ processes had a similar shape). 

Further in the future, the larger Run II data set will enable the observation and study of extremely rare SM processes. Of this family, quadruple boson production
represents a particularly interesting class of process. Large deviations from the SM could be indicative of signals from BSM physics. The observation of these attobarn processes will 
represent a significant achievement for the LHC program. Discussions are currently ongoing about the prospect of a future collider with potential center of mass energies of up to 100 TeV~\cite{Avetisyan:2013onh}. At these high energies the properties of the EW theory~\cite{Cornwall:1990hh} are such that the amplitudes for the production of $n$ Higgs and $m$ longitudinal vector bosons (at the mass threshold) 
scale as $\mathcal{A}_{1\rightarrow n+m} \sim n! m!$~\cite{Khoze:2014zha,Khoze:2014kka}. 
The production of multiple vector bosons at high energies are thus an extremely interesting class of process to study for the their perturbative behavior (or lack thereof). 
If the EW theory develops strong interactions, studies this sector would thus require serious modifications to existing theoretical predictions and procedures. 
Production of multiple photon final states will not suffer from this breakdown and thus will provide potential cross checks for studies at high energies.

In order to obtain theoretical predictions that adequately describe the experimental data, many processes will need to be extended to NNLO accuracy  
($\gamma\gamma$ already illustrates this~\cite{Catani:2011qz}). The current reach of NNLO calculations are $2\rightarrow 2$ processes. However, 
the field has experienced remarkable progress in the last couple of years ~\cite{Boughezal:2013uia,Chen:2014gva,Grazzini:2013bna,Caola:2014lpa,Gehrmann:2014fva,Cascioli:2014yka},
and as such, applications at $2\rightarrow 3$ may soon become feasible. 
Experience with $2\rightarrow2$ applications has illustrated the benefits of having stable and efficient NLO code for the $2\rightarrow 3$ processes that 
enter the real-virtual contributions to the cross section. For instance, the recent calculation of Higgs plus jet at NNLO~\cite{Boughezal:2013uia,Chen:2014gva} 
has made use of the analytic results for the one-loop $H+2j$~\cite{Dixon:2009uk,Badger:2009hw,Badger:2009vh} amplitudes. 
Therefore it is natural to assume that any future $2\rightarrow 3$ calculation will demand stable, efficient NLO code. One natural way to achieve this is 
through the means of $D$-dimensional unitarity methods~\cite{Bern:1994zx,Bern:1994cg,Britto:2004nc,Britto:2006sj,Forde:2007mi,Mastrolia:2009dr,Anastasiou:2006gt,Anastasiou:2006jv,Badger:2008cm}, which have the potential to provide 
relatively compact analytic results for $2\rightarrow 4$ one-loop basis integral coefficients.
In this paper we demonstrate the feasibility of completing $2\rightarrow 4$ calculations analytically and study their subsequent implementation into a Monte Carlo code. Given its interest as a future signal at the LHC as discussed above, and since it has not previously been studied in detail before,\footnote{A total cross section is listed in Ref.~\cite{Alwall:2014hca} and does not include photon fragmentation.} we present results for the $\gamma\gamma\gamma\gamma$ process. 
This piece is of particular interest for future $2\rightarrow 3$ NNLO calculations since it is related to 
 the most subleading in color, real-virtual, pieces of any $q\overline{q} \rightarrow m\gamma + n g$ process. In Section~\ref{sec:calc} we present a brief overview of the unitarity 
 methods and discuss several refinements that we have found to be useful in our calculation. In Section~\ref{sec:LHC} we present a phenomenological 
 study of this process at the LHC, and extensions to future high energy machines. In Section~\ref{sec:conc} we present our conclusions.

\section{Calculation} 
\label{sec:calc}

In this section we describe the details of the calculation of the process $pp\rightarrow \gamma\gamma\gamma\gamma$ at Next-to-Leading Order. 
The LO amplitude for the production of four-photon final states can be written as follows,
\begin{eqnarray}
\mathcal{A}_6^{(0)}(1_{q}^{h_1},2_{\overline{q}}^{h_2},3_{\gamma}^{h_3},4_{\gamma}^{h_4},5_{\gamma}^{h_5},6_{\gamma}^{h_6})=i(\sqrt{2}e Q_i)^4 A^{(0)}_6(1_{q}^{h_1},2_{\overline{q}}^{h_2},3_{\gamma}^{h_3},4_{\gamma}^{h_4},5_{\gamma}^{h_5},6_{\gamma}^{h_6}) \,.
\end{eqnarray} 
At NLO we must consider one-loop (virtual) and real amplitudes, the virtual amplitudes have a similar decomposition to those at LO, 
\begin{eqnarray}
\mathcal{A}_6^{(1)}(1_{q}^{h_1},2_{\overline{q}}^{h_2},3_{\gamma}^{h_3},4_{\gamma}^{h_4},5_{\gamma}^{h_5},6_{\gamma}^{h_6})=\frac{\alpha_S}{2\pi}\left(\frac{N_c^2-1}{N_c}\right)i(\sqrt{2}e Q_i)^4 A^{(1)}_6(1_{q}^{h_1},2_{\overline{q}}^{h_2},3_{\gamma}^{h_3},4_{\gamma}^{h_4},5_{\gamma}^{h_5},6_{\gamma}^{h_6}) \,.
\end{eqnarray} 
Finally the real radiation terms can be constructed from the following amplitude, 
\begin{eqnarray}
\mathcal{A}_7^{(0)}(1_{q}^{h_1},2_{\overline{q}}^{h_2},3_{\gamma}^{h_3},4_{\gamma}^{h_4},5_{\gamma}^{h_5},6_{\gamma}^{h_6},7_g^{h_7})=\sqrt{2}g_s (T^{a_7}_{i_1i_2})i(\sqrt{2}e Q_i)^4 A^{(0)}_7(1_{q}^{h_1},2_{\overline{q}}^{h_2},3_{\gamma}^{h_3},4_{\gamma}^{h_4},5_{\gamma}^{h_5},6_{\gamma}^{h_6},7_g^{h_7}) \,.
\end{eqnarray}
Helicity amplitudes for the tree-level processes $\mathcal{A}_6^{(0)}$ and $\mathcal{A}_7^{(0)}$ are straightforward to compute. Therefore we focus in this section 
on the the calculation of the one-loop amplitudes  $\mathcal{A}_6^{(1)}$, which are significantly more involved. 
We note that we only need to calculate two one-loop scattering amplitudes\textemdash{}the MHV and the NMHV amplitudes for $q\overline{q}\rightarrow \gamma\gamma\gamma\gamma$:
\begin{align}
\label{helamp}
A^{(1)}_6(1_{q}^{-},2_{\overline{q}}^{+},3_{\gamma}^{-},4_{\gamma}^{+},5_{\gamma}^{+},6_{\gamma}^{+}),&& 
A^{(1)}_6(1_{q}^{-},2_{\overline{q}}^{+},3_{\gamma}^{-},4_{\gamma}^{-},5_{\gamma}^{+},6_{\gamma}^{+}) \,.
\end{align}
All other helicity configurations can be obtained from these by trivial relabeling and chiral conjugation. 
To calculate these two amplitudes, we use the $D$-dimensional unitarity method \cite{Bern:1994zx,Bern:1994cg,Britto:2004nc,Britto:2006sj,Forde:2007mi,Mastrolia:2009dr,Anastasiou:2006gt,Anastasiou:2006jv,Badger:2008cm}, wherein one evaluates unitarity cuts in $D=4-2\epsilon$ dimensions and performs a series expansion in the square of the $(-2\epsilon)$-dimensional components of the loop momentum $\ell$, i.e. a series expansion in large $\mu^2$ where $\ell^2 = \ell_{(4)}^2 + \mu^2$. This gives a decomposition of each one-loop amplitude into a linear combination of scalar bubble ($I_2$), triangle ($I_3$), and box ($I_4$) integrals plus additional rational terms ($R_n$) arising from higher-order in $\mu^2$ terms:
\begin{align}
A_n^{(1)} &= \sum_{K_1,K_2,K_3,K_4} d_{K_i} I_{4}(K_i;\epsilon) + \sum_{K_1,K_2,K_3} c_{K_i} I_3(K_i;\epsilon) + \sum_{K} b_{K} I_2(K;\epsilon)  + R_n + \mathcal{O}(\epsilon) \,,
\end{align}
where $K_i$ specify the momenta entering the corners of the bubble, triangle, and box integrals. The rational terms can be further decomposed into contributions from higher-order in $\mu^2$ box, triangle, and bubble integrals as
\begin{align}
R_n &= -\frac{1}{6}  \sum_{K_1,K_2,K_3,K_4} d_{K_i}^{[4]} - \frac{1}{2} \sum_{K_1,K_2,K_3} c_{K_i}^{[2]} - \frac{1}{6} \sum_{K}(K^2 - 3(m_1^2 +m_2^2)) b_{K}^{[2]},
\end{align}
where $m_1$ and $m_2$ are the masses of the internal lines. The numbers in square brackets indicate the degree in $\mu$. 

For the particular amplitudes in (\ref{helamp}), the box and bubble contributions to the rational terms ($d_{K_i}^{[4]}$ and $b_{K}^{[2]}$) all vanish. Techniques for calculating the remaining coefficients $b_{K}$, $c_{K_i}$, $d_{K_i}$, and $c_{K_i}^{[2]}$ are well-known and mature; in this section, we do not attempt to give a complete exposition of them, but rather we briefly present some modifications that we have found to be convenient, in particular for the three-mass triangle coefficients $c_{K_i}$.

\subsection{Triangle Coefficients}
\begin{figure}
\begin{center}
	\includegraphics[scale=0.7]{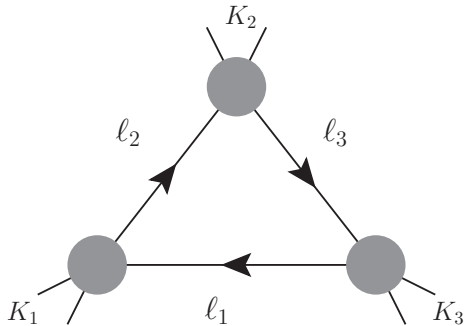}
\end{center}
	\caption{Momentum flow and labelling for the triangle cuts used in the text. External momenta are taken as incoming.}
	\label{tricutfig}
\end{figure}
Much of the complexity of analytic $2\rightarrow4$ NLO calculations is caused by the presence of three-mass triangle integrals, for which a schematic representation is illustrated in Fig.~\ref{tricutfig}. These contributions also pollute the bubble integral coefficients and the rational terms and make results significantly lengthier than for amplitudes that lack three-mass triangles. Three-mass triangles already appear in $2\rightarrow2$ and $2\rightarrow3$ amplitudes with massive external states such as $W/Z$, and their impact can be observed for instance by comparing the complexity of 
$q\overline{q}\rightarrow ggg$~\cite{Bern:1994fz} amplitudes to those for $q\overline{q}\rightarrow Zgg$~\cite{Bern:1997sc}. At the $2\rightarrow4$ level these pieces occur frequently; looking forward to the next wave of NNLO precision calculations, it will be very helpful to develop an analytic finesse for the calculation of these coefficients. In this section we present a method that we have found to be efficient in this regard.

One standard method for calculating the triangle coefficients $c_{K_i}$ comes from Forde~\cite{Forde:2007mi}. In this method, one solves the three triangle cut constraints ($\ell_1^2 = 0$, $\ell_2^2 = (\ell_1+K_1)^2 = 0$, $\ell_3^2 = (\ell_1-K_3)^2 = 0$) in terms of an affine variable $t$ and the quantities
\begin{align}
K_{ij}^{\flat\mu} &= \frac{1}{2\sqrt{\Delta}} \left( \gamma_{ij} K_i^\mu - K_i^2 K_j^\mu \right), \\
\gamma_{ij} &= K_i\cdot K_j +\sqrt{\Delta}, \\
\overline{\gamma}_{ij} &= K_i\cdot K_j -\sqrt{\Delta}, \\
\Delta &= (K_i\cdot K_j)^2 - K_i^2 K_j^2.
\end{align}
The flatted momenta here are null linear combinations of the momenta flowing into the corners of the triangle. They are also very useful for expressing bubble coefficients and rational terms.
$\Delta$ is equal to the negative of the triangle Gram determinant and does not depend on the choice of $i$ and $j$. Ref.~\cite{Forde:2007mi} gives a carefully chosen parameterization of the cut momenta as
\begin{align}
\label{at1}
|\ell_1\rangle &= t |K_{13}^\flat\rangle + \frac{\gamma_{12}}{2\sqrt{\Delta}} |K_{31}^\flat\rangle, &
|\ell_1] &= -\frac{\gamma_{23}}{2t \sqrt{\Delta}} |K_{13}^\flat]   +   |K_{31}^\flat], \\
\label{at2}
|\ell_2\rangle &= t |K_{13}^\flat\rangle + \frac{\overline{\gamma}_{13} \gamma_{23}}{2\sqrt{\Delta}K_3^2} |K_{31}^\flat\rangle, &
|\ell_2] &= -\frac{\overline{\gamma}_{23}}{2t\sqrt{\Delta}} |K_{13}^\flat] + |K_{31}^\flat], \\
\label{at3}
|\ell_3\rangle &= t|K_{13}^\flat\rangle -\frac{\overline{\gamma}_{12}}{2\sqrt{\Delta}} |K_{31}^\flat\rangle, &
|\ell_3] &= -\frac{\overline{\gamma}_{13}\gamma_{12}}{2t\sqrt{\Delta} K_1^2}|K_{13}^\flat] + |K_{31}^\flat],
\end{align}
where we have adjusted for our labelling conventions, which are shown in Fig.~\ref{tricutfig}. With this parameterization, the triangle cut integration measure becomes
\begin{align}
\int \Diff{4} \ell \prod_{i=1}^3 \delta(\ell_i^2) &\rightarrow \int \frac{\diff t}{t \overline{\gamma}_{13}},
\end{align}
and the triangle coefficient $c_{K_1|K_2|K_3}$ can be isolated by expanding the product of three tree amplitudes $A_1(t) A_2(t) A_3(t)$ around $t\rightarrow \infty$ and selecting the $t^0$ term:
\begin{align}
c_{K_1|K_2|K_3} &= -[\text{Inf}_tA_1A_2A_3](t)\bigr|_{t=0} \,.
\end{align}

An alternative approach alters the Feynman $i\epsilon$ prescription to express the triangle coefficient as a difference of two bubble coefficients \cite{Mastrolia:2006ki}. This has the benefit of allowing the use of powerful spinor integration methods that come out of the bubble analysis. In our calculations, we find it convenient to take a hybrid approach similar in spirit to Ref.~\cite{BjerrumBohr:2007vu}; we use the following parameterization for the triangle loop momentum in terms of a homogeneous $\mathbb{C}P^1$ variable $\lambda$:
\begin{align}
\label{ht1}
|\ell_1\rangle &= |\lambda\rangle,  &
|\ell_1] &= -\frac{K_1K_2K_3|\lambda\rangle}{\Spaa{\lambda|K_1K_2|\lambda}}, \\
\label{ht2}
|\ell_2\rangle &= K_2K_3|\lambda\rangle,    &
|\ell_2] &= -\frac{K_1|\lambda\rangle}{\Spaa{\lambda|K_1K_2|\lambda}},    \\
\label{ht3}
|\ell_3\rangle &= K_2K_1|\lambda\rangle, &
|\ell_3] &= -\frac{K_3|\lambda\rangle}{\Spaa{\lambda|K_1K_2|\lambda}}.
\end{align}
Although the mass dimensions of these spinors are not traditional, the scaling properties of the amplitudes ensure that only the combinations $[\ell_i|\gamma^\mu|\ell_i\rangle$ 
matter in the cuts. With this parameterization, the triangle cut integration  becomes

\begin{align}
\int \Diff{4} \ell \prod_{i=1}^3 \delta(\ell_i^2) &\rightarrow \oint \frac{\Spaa{\lambda \diff\lambda}}{\Spaa{\lambda|K_1K_2|\lambda}}.
\end{align}
\begin{figure}
\begin{center}
	\includegraphics[scale=0.6]{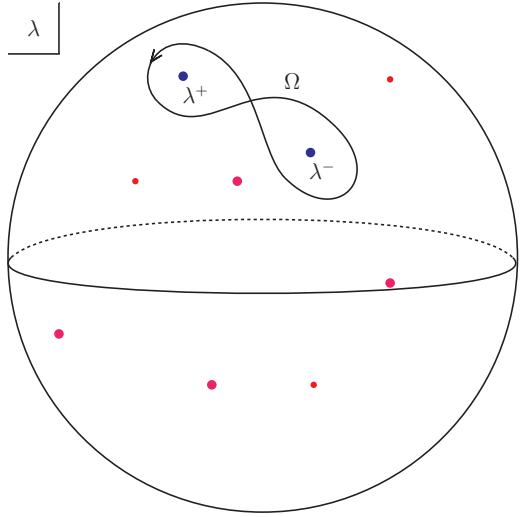}
\end{center}
	\caption{The contour $\Omega$ on the Riemann sphere of $\lambda$ is chosen to isolate the two poles of the triangle integration measure, $\lambda^\pm$ (shown in blue), from the box poles (shown in red). The sum of all of the residues vanishes.}
	\label{contourfig}
\end{figure}
We do not need to know the physical integration contour. Instead, we deform to an unphysical contour that isolates the triangle coefficient and discards box contributions. That is, we select a contour $\Omega$ such that  
\begin{align}
\label{eqtricutcontour}
\oint_\Omega  \frac{\Spaa{\lambda \diff\lambda}}{\Spaa{\lambda|K_1K_2|\lambda}} A_1(\lambda)A_2(\lambda)A_3(\lambda) &=
 	\oint_\Omega \frac{\Spaa{\lambda \diff\lambda} }{\Spaa{\lambda|K_1K_2|\lambda}} c_{K_1|K_2|K_3}.
\end{align}
To get a nonzero result on the right hand side, $\Omega$ should encircle at least one of the two poles at $\Spaa{\lambda|K_1K_2|\lambda}=0$, but not both in the same direction. Let us call these two zeros $\lambda^+$ and $\lambda^-$. On the left hand side, box contributions arise from poles in the tree-level amplitudes $A_i(\lambda)$. The sum of the residues at $\lambda^+$ and $\lambda^-$ is equal to negative the sum of all of the box residues, and the only other combination available to us is a contour that encircles $\lambda^+$ and $\lambda^-$ in opposite directions. This contour choice is illustrated in Fig.~\ref{contourfig}.

Unsurprisingly, $|\lambda^+\rangle$ and $|\lambda^-\rangle$ are precisely $|K_{12}^\flat\rangle$ and $|K_{21}^\flat\rangle$ (also projectively equivalent to $|K_{31}^\flat\rangle$ and $|K_{13}^\flat\rangle$, or $|K_{23}^\flat\rangle$ and $|K_{32}^\flat\rangle$). This can be seen by factorizing
\begin{align}
	\Spaa{\lambda|K_1K_2|\lambda} &= \frac{2 \sqrt{\Delta}  }{\Spaa{K_{12}^\flat K_{21}^\flat}}\Spaa{\lambda K_{12}^\flat} \Spaa{\lambda K_{21}^\flat}.
\end{align}
Alternatively, it is possible to avoid flatted momenta entirely by observing that the matrix 
\begin{align}
\label{M2}
M &= \frac{1}{2}(K_1K_2-K_2K_1)
\end{align}
is a projective involution: $M^2 = \Delta$. Then we can solve $\Spaa{\lambda^\pm|K_1K_2|\lambda^\pm}=0$ by introducing a generic auxiliary parameter $\eta$ and taking
\begin{align}
|\lambda^\pm\rangle &= \left(\frac{M}{\sqrt{\Delta}}\pm1\right)|\eta\rangle.
\end{align}
In this case,
\begin{align}
 \Spaa{\lambda|K_1K_2|\lambda} &= \frac{\Spaa{\lambda|(M{+}\sqrt{\Delta})|\eta}\, \Spaa{\lambda|(M{-}\sqrt{\Delta})|\eta}}{\Spaa{\eta|M|\eta}}.
\end{align}
Regardless of the approach one wishes to take, the integral on the right hand side of eq.~(\ref{eqtricutcontour}) evaluates to
\begin{align}
\oint_\Omega \frac{\Spaa{\lambda \diff\lambda} }{\Spaa{\lambda|K_1K_2|\lambda}} c_{K_1|K_2|K_3} &= 
	\frac{2\pi i}{\sqrt{\Delta}} c_{K_1|K_2|K_3}.
\end{align}
Thus, one can isolate the coefficient of the triangle integral by calculating
\begin{align}
c_{K_1|K_2|K_3} &= (\text{Res}_{\lambda^+}-\text{Res}_{\lambda^-})\left[ 
	\frac{\sqrt{\Delta}}{\Spaa{\lambda|K_1K_2|\lambda}} A_1(\lambda)A_2(\lambda)A_3(\lambda)
	\right] \,.
\end{align}

Let us now return to the four-photon process at hand. The MHV amplitude $A^{(1)}_6(1_{q}^{-},2_{\overline{q}}^{+},3_{\gamma}^{-},4_{\gamma}^{+},5_{\gamma}^{+},6_{\gamma}^{+})$ has no three-mass triangle contributions and is relatively straightforward to calculate. The NMHV amplitude $A^{(1)}_6(1_{q}^{-},2_{\overline{q}}^{+},3_{\gamma}^{-},4_{\gamma}^{-},5_{\gamma}^{+},6_{\gamma}^{+})$ has six non-vanishing three-mass triangle coefficients, which we can label 
\begin{align}
& c_{16|35|24},\quad c_{15|36|24},\quad c_{16|45|23},\quad c_{15|46|23},\quad c_{12|35|46},\quad c_{12|36|45}. 
\end{align}
All but two of these are related by trivial symmetries of the amplitude.  The remaining two are given by
\begin{align}
c_{12|35|46} &=  (\text{Res}_{\lambda^+}-\text{Res}_{\lambda^-})\left[ 
	\frac{2 \Spaa{1\lambda}^2 \Spaa{3|K_2K_3|\lambda}^2 \Spaa{4|K_2K_1|\lambda}^2 \sqrt{\Delta}}
		{ \Spaa{2\lambda} \Spaa{6\lambda} \Spaa{1|K_2K_3|\lambda} \Spaa{5|K_2K_1|\lambda} \Spaa{5|K_2K_3|\lambda} \Spaa{6|K_2K_1|\lambda} \Spaa{\lambda|K_1K_2|\lambda} }
	\right], \\
c_{16|35|24} &=  (\text{Res}_{\lambda^+}-\text{Res}_{\lambda^-})\left[ 
 	\frac{\Spaa{1\lambda}^2 \Spaa{3|K_2K_3|\lambda}^2 \Spaa{4|K_2K_1|\lambda}^2  \sqrt{\Delta}  }{ \Spaa{16} \Spaa{2\lambda} \Spaa{5|K_2K_3|\lambda} \Spaa{5|K_2K_1|\lambda} \Spaa{6|K_2K_3|\lambda} \Spaa{\lambda|K_1K_2|\lambda}^2  } 
	\right],
\end{align}
where $\Delta$ and $K_i$ are those appropriate to each triangle. To take the residues, one can perform a partial fraction decomposition on $\lambda$ using Schouten identities and then apply the basic identities
\begin{align}
 (\text{Res}_{\lambda^+}-\text{Res}_{\lambda^-})\left[ 
	\frac{\sqrt{\Delta} \Spaa{a\lambda}}{\Spaa{b\lambda}\Spaa{\lambda|K_1K_2|\lambda}}
	\right] &= \frac{\Spaa{a|[K_1,K_2]|b}}{2\Spaa{b|K_1K_2|b}},  \\
 (\text{Res}_{\lambda^+}-\text{Res}_{\lambda^-})\left[ 
	\frac{\sqrt{\Delta} \Spaa{a_1\lambda}\Spaa{a_2\lambda}}{\Spaa{\lambda|K_1K_2|\lambda}^2}
	\right] &= -\frac{1}{4\Delta} \Spaa{a_1|[K_1,K_2]|a_2}, \\
 (\text{Res}_{\lambda^+}-\text{Res}_{\lambda^-})\left[ 
	\frac{\sqrt{\Delta} \lambda^{\alpha_1}\cdots \lambda^{\alpha_{2m}} } {\Spaa{\lambda|K_1K_2|\lambda}^{m+1}}
	\right] &= \frac{1}{(m!)^2 (8\Delta)^m} [K_1,K_2]^{(\alpha_1\alpha_2} \cdots [K_1,K_2]^{\alpha_{2m-1}\alpha_{2m})} \,.
\end{align}
There are $(2m-1)!!$ different terms in the symmetrization over $\alpha_i$ on the RHS, which, together with the number of box poles, provides a rough upper limit to the complexity one can expect from the answer. By taking the residues in this way, all of the spurious Gram determinant poles can be made manifest algorithmically. In order to avoid introducing a host of other unphysical poles, it is helpful to limit oneself to the partial fraction identities
\begin{align}
 \frac{\Spaa{a_1\lambda}\Spaa{a_2\lambda}}{\Spaa{b\lambda}\Spaa{\lambda|K_1K_2|\lambda}} &= 
 	\frac{\Spaa{a_1b}\Spaa{a_2b}}{\Spaa{b|K_1K_2|b}} \frac{1}{\Spaa{b\lambda}} +
	\frac{\Spaa{b|K_1K_2|a_1}\Spaa{a_2\lambda}-\Spaa{b a_1}\Spaa{a_2|K_1K_2|\lambda}  }{\Spaa{b|K_1K_2|b}} \frac{1}{\Spaa{\lambda|K_1K_2|\lambda}}, \\
\frac{\Spaa{\lambda|M_1|\lambda}}{\Spaa{b\lambda}\Spaa{\lambda|M_2|\lambda}} &=
	\frac{\Spaa{b|M_1|b}}{\Spaa{b|M_2|b}} \frac{1}{\Spaa{b\lambda}} +
	\frac{\Spaa{b|[M_1,M_2]|\lambda} }{\Spaa{b|M_2|b}} \frac{1}{\Spaa{\lambda|M_2|\lambda}},
\end{align}
where $M_1$ and $M_2$ are any matrices (with the correct Lorentz indices) sandwiched inside the spinor products. The only poles that are introduced occur when $\Spaa{b|K_1K_2|b}=0$, which is precisely when a box pole collides with one of the $\lambda^{\pm}$ poles and pinches the contour $\Omega$. Thus the limit $\Spaa{b|K_1K_2|b}\rightarrow 0$ does not commute with the contour integration, but rather $c_{K_1|K_2|K_3}$ picks up a box contribution in this limit.

One can perform the same procedure on triangles with massless legs. The parameterization in eqs.~(\ref{ht1}) -- (\ref{ht3}) reduces straightforwardly in the limit where any of the $K_i$ are massless, with the caveat that one of the two residues at $\lambda^\pm$ will always vanish. To get the second residue, one can use either the chiral conjugate or a cyclic permutation of the parameterization (\ref{ht1}) -- (\ref{ht3}).

\subsection{Box Cuts}

Although four-mass box cuts are not relevant in four-photon production, our triangle parameterization suggests a convenient way to deal with them. We include it here for completeness. Consider cutting a fourth leg on the $K_3$ corner of a triangle cut, so that $K_3 \rightarrow K_3+K_4$ splits into two corners of a box cut. Then, using the parameterization in eq.~(\ref{ht1}), the final cut constraint $2\ell_1\cdot K_4 = K_4^2$ is
\begin{align}
\frac{\Spaa{\lambda|K_1K_2(K_3+K_4)K_4|\lambda}}{\Spaa{\lambda|K_1K_2|\lambda}} &= K_4^2,
\end{align}
which implies
\begin{align}
\frac{\Spaa{\lambda|K_1K_2K_3K_4|\lambda}}{\Spaa{\lambda|K_1K_2|\lambda}} &= 0.
\end{align}
Just as for the matrix in eq.~(\ref{M2}), we observe that the matrix 
\begin{align}
M_4 &= \frac{1}{2}(K_1K_2K_3K_4 - K_4K_3K_2K_1)
\end{align}
is a projective involution:
\begin{align}
M_4^2 &= \Delta_4 = \frac{1}{16} \text{tr}(K_1K_2K_3K_4)^2 - K_1^2 K_2^2 K_3^2 K_4^2 .
\end{align}
This is (unsurprisingly) proportional to the quantity that appears in a square root in $\rho$ in the solution of Ref.~\cite{Britto:2004nc}.
Then using a generic auxiliary $\mathbb{C}P^1$ variable $\eta$, the two solutions to the constraint are
\begin{align}
 |\lambda^\pm \rangle &= \left(\frac{M_4}{\sqrt{\Delta_4}} \pm 1\right)|\eta\rangle .
\end{align}
These two solutions can be inserted back into the parameterization in eqs.~(\ref{ht1}) -- (\ref{ht3}) to give the two box cut solutions. In the degenerate case where at least one of the corners is massless, this immediately reproduces the form of the box loop momentum given in Refs.~\cite{Risager:2008yz} and \cite{Berger:2008sj}.

\subsection{Rational Terms}
For the rational terms $R_n$, we use the $D$-dimensional unitarity techniques of Ref.~\cite{Badger:2008cm}. In the amplitudes at hand, the only contributions to the rational terms come from $D+2$-dimensional triangle integrals. Since all of the cuts we need are leading in $\mu^2$, we can use the following loop momentum parameterization:
\begin{align}
	\ell_1^\nu &= \frac{\mu}{\sqrt{\gamma_{13} (2K_{13}^\flat\cdot \eta)(2K_{31}^\flat \cdot \eta)}} \left( t\, \Spab{\eta|K_{13}^\flat \gamma^\nu K_{31}^\flat|\eta}- \frac{1}{t}\, \Spab{\eta|K_{31}^\flat \gamma^\nu K_{13}^\flat|\eta}\right),
\end{align}
with $\eta$ a generic null reference vector. Following Ref.~\cite{Badger:2008cm}, we extract the $D+2$-dimensional triangle coefficient as
\begin{align}
c^{[2]} &= \frac{1}{2}\sum_\pm \text{Inf}_{\mu^2} \left[ \text{Inf}_t[A_1A_2A_3(\ell^\pm)]|_{t^0} \right]|_{\mu^2},
\end{align}
where the sum is over the two choices of sign on $\sqrt{\Delta}$ in the loop momentum parameterization. With the parameterization used here, flipping the sign on $\sqrt{\Delta}$ is equivalent to sending $t\rightarrow \frac{1}{t}$, and it is possible to interchange the order of limits $t\rightarrow \infty$ and $\mu^2 \rightarrow \infty$. In this way, we calculate 
\begin{align}
c^{[2]} &= \frac{1}{2} \text{Inf}_{\mu^2} [ A_1 A_2 A_3(\ell)]|_{\mu^2} \left(|_{t^0} + |_{(\frac{1}{t})^0} \right).
\end{align}
Since $\mu^2$ is the leading behaviour in all of the needed cuts, the $\mu^2$ limit is immediate.

\subsection{Implementation}

The analytic calculations in this paper were performed in Mathematica, and we have made use of the S@M package~\cite{Maitre:2007jq}. 
LO and real amplitudes have been checked against Madgraph~\cite{Alwall:2011uj} and an in-house numerical computation. The virtual 
amplitudes have been checked against an implementation of the numerical $D$-dimensional unitarity algorithm presented in Ref.~\cite{Ellis:2008ir}.
We have implemented all of the unitarity cuts, including rational terms, for $q\overline{q}\rightarrow\gamma\gamma\gamma\gamma$ analytically into MCFM \cite{Campbell:1999ah,Campbell:2011bn,MCFMweb}. The formulae are too long to include in full in this publication, but may be inspected in the distributed code. The resulting code runs stably in double precision and will be released publicly in a forthcoming version of MCFM.

\section{Phenomenology}
\label{sec:LHC}

\subsection{$\gamma\gamma\gamma\gamma$ at the LHC}

In this section we investigate the phenomenology of four photon final states at the LHC. Since the cross section 
is rather small, we focus primarily on the LHC operating at Run II energies of 13 and 14 TeV. We use the default EW parameters of MCFM \cite{Campbell:1999ah,Campbell:2011bn,MCFMweb},
of which $\alpha=1/132.338$ is the most relevant for the discussion here. We apply the following phase space selection cuts to the photons: 
\begin{eqnarray} 
p_{T}^{\gamma} > 20 \;{\rm {GeV}}, \quad  |\eta_{\gamma}|  < 2.5, \quad R_{\gamma\gamma} > 0.4 \,.
\end{eqnarray} 
In addition to the above requirements, photons are isolated by requiring that the hadronic energy inside a cone $R_0 =0.4$ about the photon is restricted to be less than a chosen fraction of the photon's energy
\begin{eqnarray} 
\sum_{i \in R_0} E_{T,i}^{had} \le \epsilon_{\gamma} p_{T}^{\gamma} \,.
\label{eq:fraciso}
\end{eqnarray}
We set $\epsilon_{\gamma} = 0.5$ as our default value. The above isolation condition requires the inclusion of photon fragmentation functions~\cite{Bourhis:1997yu,GehrmannDeRidder:1997gf},\footnote{See e.g. Ref.~\cite{Campbell:2014yka} and references therein for a more detailed discussion of photon isolation effects at the LHC.} for which we use the LO set of GdRG~\cite{GehrmannDeRidder:1997gf}.
For NLO (LO) calculations we use the CT10~\cite{Lai:2010vv}  (CTEQ6L1) PDF sets. Unless otherwise stated our default renormalziation, factorization and fragmentation scales are set to $\mu=m_{\gamma\gamma\gamma\gamma}$.  Infrared singularities are regulated using the Catani Seymour dipole approach~\cite{Catani:1996vz,Catani:2002ny}.
\begin{figure}
\begin{center}
\includegraphics[width=8cm]{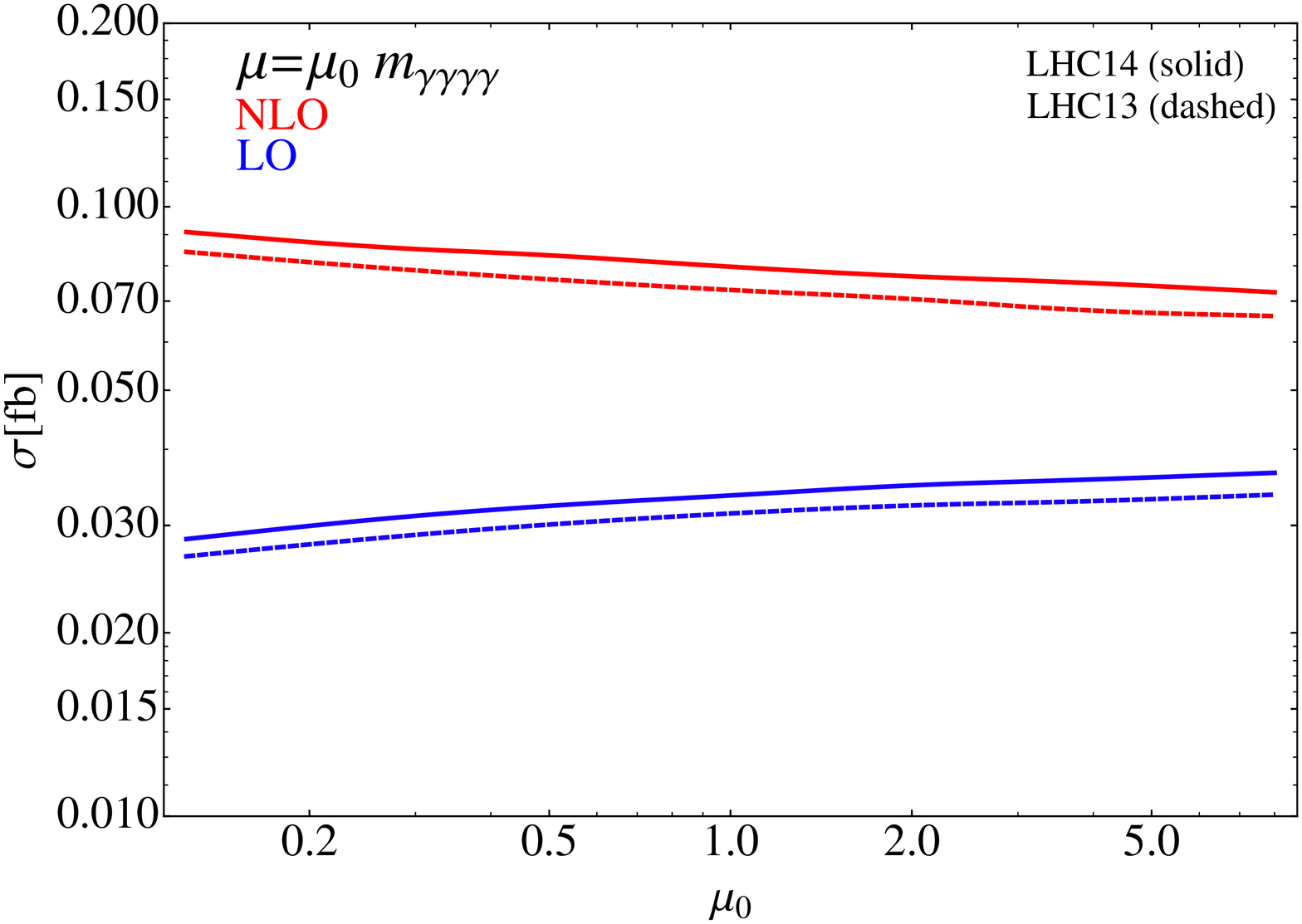}
\caption{Scale dependence of the LO and NLO cross sections using the basic phase space cuts at the LHC14 (solid) and LHC13 (dashed).  The 
central scale choice is set to the invariant mass of the four-photon final state $\mu=m_{\gamma\gamma\gamma\gamma}$, and excursions around this choice 
by a factor of $\mu_0$ are illustrated.}
\label{fig:scaldep}
\end{center}
\end{figure}

In Fig.~\ref{fig:scaldep} we present the dependence of the LO and NLO cross section on the renormalization and factorization scale $\mu=\mu_0m_{\gamma\gamma\gamma\gamma}$. 
From the curve it is clear that this process possesses large $K$-factors, for $\mu_0=1$, the $K$-factor for this process is 2.37 (at 14 TeV), whilst at the values $\mu_0=1/4$ (4) the  $K$-factor 
increases (decreases) to 3.19 (1.98). The scale dependence at LO is very weak, with a small growth in the cross section as the scale increases. 
For the NLO prediction, the cross section decreases as the scale increases. All of these features are completely consistent with those observed in similar processes.\footnote{See e.g. the 
discussion in Ref.~\cite{Campbell:2011bn}.} At LO there is essentially no scale dependence; since this is a EW process, no $\alpha_S$ enters the LO matrix element. Indeed, 
the only source of scale dependence comes from the factorization scale of the PDFs. The dependence on $\alpha_S$ explicitly enters the calculation at NLO and results in 
a partial cancellation between the renormalization scale (which suppresses the cross section at large $\mu_0$) and the factorization scale  (which increases the cross section at large $\mu_0$). 
The net result is a rather small dependence on the scale choice, e.g. $\sigma(\mu=0.5 m_{4\gamma})/\sigma(\mu=2m_{4\gamma})=1.04$. As a result, care must be taken if
interpreting this range as a legitimate theoretical uncertainty estimate. A more sensible choice is to widen the range to $\mu_0=1/4$ and $\mu_0=4$, which results in a $\sim \pm 8\%$ uncertainty. However, given the large $K$-factor for this process, this choice is still unlikely to incorporate the NNLO corrections in the range. 

\begin{figure}
\begin{center}
\includegraphics[width=8cm]{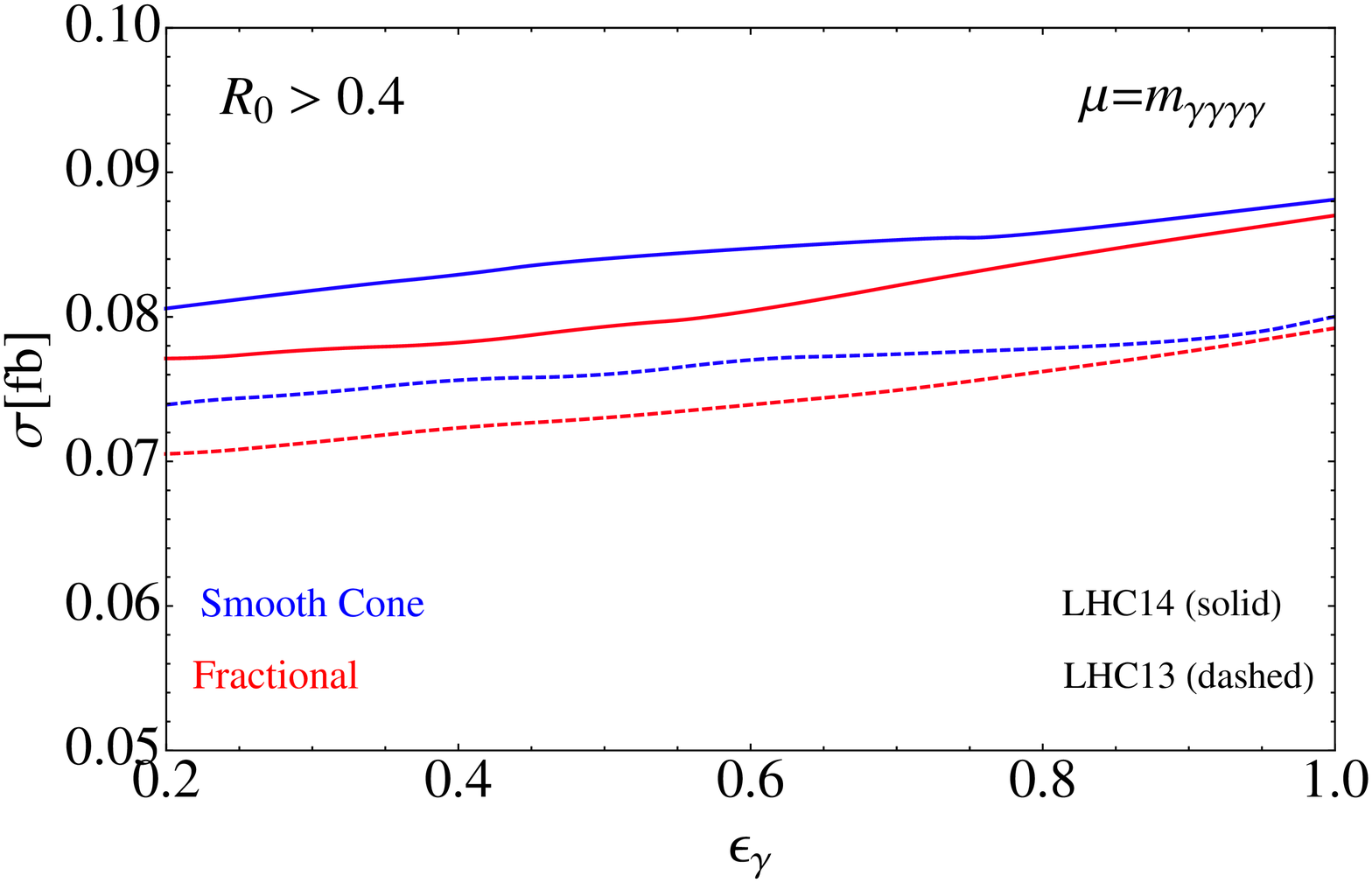}\
\includegraphics[width=8cm]{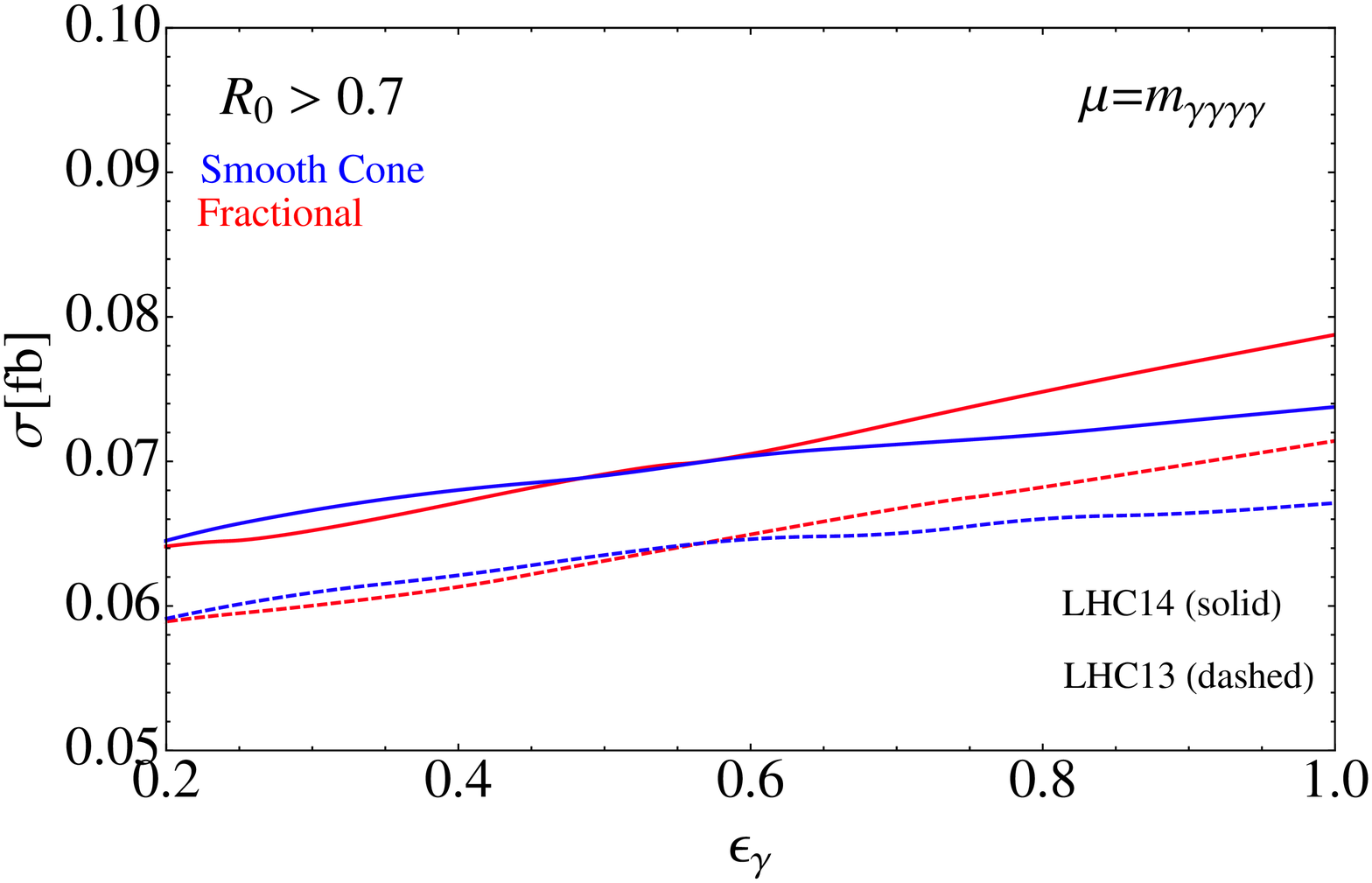}
\caption{Dependence of the NLO cross section for different isolation criteria at 13 and 14 TeV. The plot on the left hand side corresponds to a cone size 
of $R_{0} > 0.4$, whilst the plot on the right illustrates the dependence for the larger cone size choice of $R_{0} > 0.7$.}
\label{fig:isodep}
\end{center}
\end{figure}

Figure~\ref{fig:isodep} shows the dependence of the NLO cross section on the two different isolation requirements. The red curves correspond to the isolation 
prescription defined in eq.~(\ref{eq:fraciso}) (``fractional isolation"), while the blue curves correspond to the procedure presented in Ref.~\cite{Frixione:1998jh}, defined here as ``smooth cone"
isolation. The smooth cone isolation criterion requires that the hadronic energy in the vicinity of the photon satisfies the following requirement, 
\begin{eqnarray}
\sum_{{\rm{had}}} E_T^{{\rm{had}}}\theta(R-R_{{\rm{had}},\gamma}) < \epsilon_\gamma p_T^{\gamma} \left(\frac{1-\cos{R}}{1-\cos{R_0}}\right)^n \quad \forall \; R \le R_0 \,.
\end{eqnarray}
In the above equation $R_0$ and $n$ are input parameters of the isolation. In our example we choose $R_0$ democratically for smooth cone and fractional 
isolation requirements and plot the resulting cross section as a function of $\epsilon_{\gamma}$, treating the parameter equivalently in both prescriptions. 
Our results illustrate that, over the range of $\epsilon_{\gamma}$ studied above, the cross section obtained with the fractional isolation is broadly compatible with that obtained 
using the smooth cone isolation. The largest deviations occur for large cone sizes $R_0 > 0.7$ and loose isolation $\epsilon_{\gamma} =1$. The shapes of the curves are 
insensitive to the choice of operating energies. Over the bulk of the range, the cross sections from the two types of isolation lie within each other's scale uncertainty, particularly 
if the central scale is varied by a factor of four in each direction, resulting in an $\mathcal{O}(10\%)$ band for each curve. 
 
It is interesting to compare Fig.~\ref{fig:isodep} with the similar results obtained for the triphoton (and $\gamma\gamma+$jet) process presented in Ref.~\cite{Campbell:2014yka}.
The results of Ref.~\cite{Campbell:2014yka} showed similar results (at 14 TeV) for the differences between fractional and smooth cone isolation. Ref.~\cite{Campbell:2014yka} also illustrated that the shape of the curves as a function of $\epsilon_{\gamma}$ were very similar for the $\gamma\gamma\gamma$ and $\gamma\gamma$+jet final states. The latter observation indicates that the isolation dependence is more sensitive to the underlying $2\rightarrow 4$ kinematics of the real phase space than to the number of final state photons. Comparing the results for triphotons with those for four photons presented here, we observe that the shapes of the two curves remain broadly the same. Therefore it is tempting to suggest that the results used above can be used 
to estimate the choice of smooth cone parameters in cases such as $\gamma\gamma+2$ jets, for which existing predictions do not include fragmentation. 

\begin{figure}
\begin{center}
\includegraphics[width=4cm]{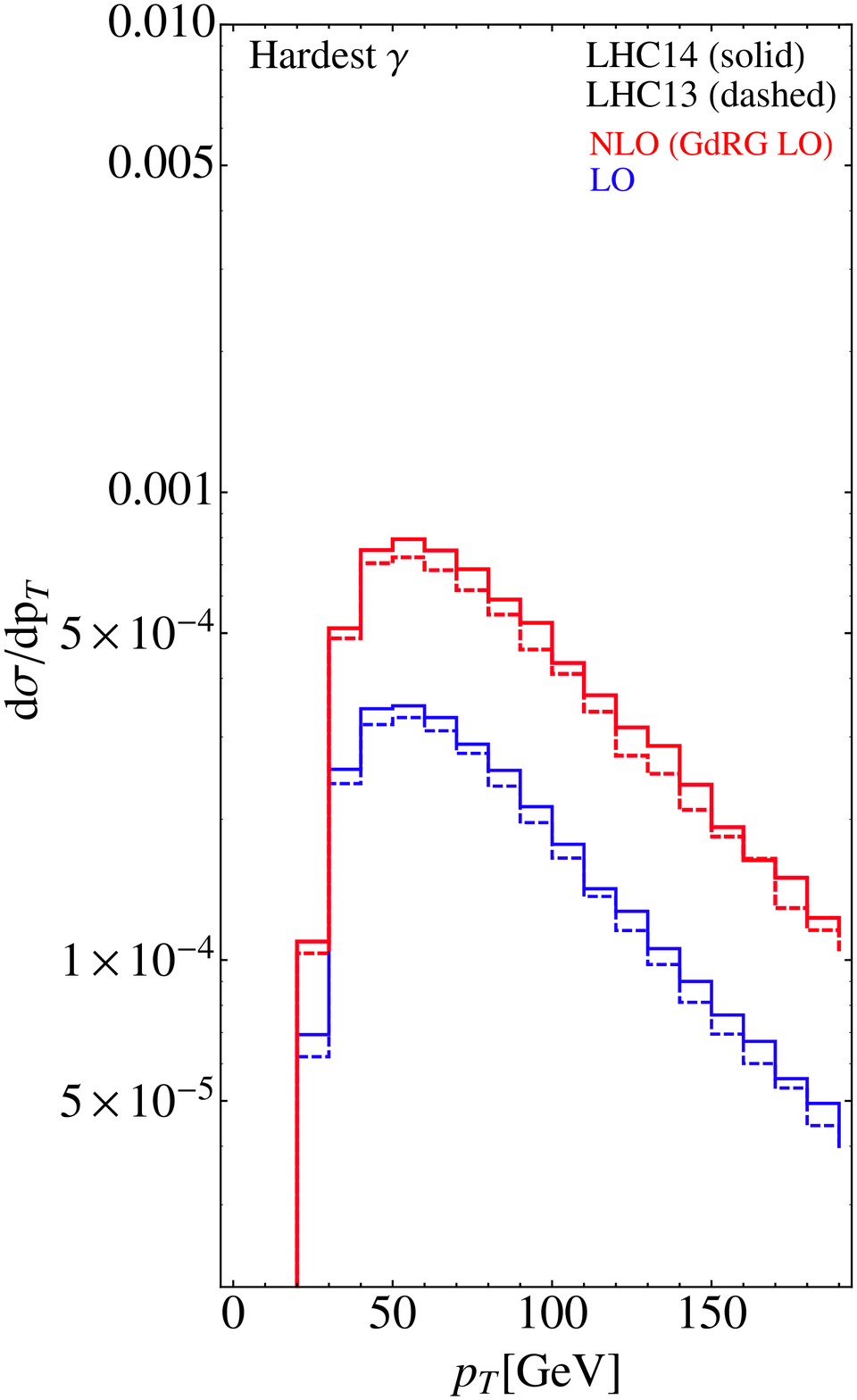}
\includegraphics[width=4cm]{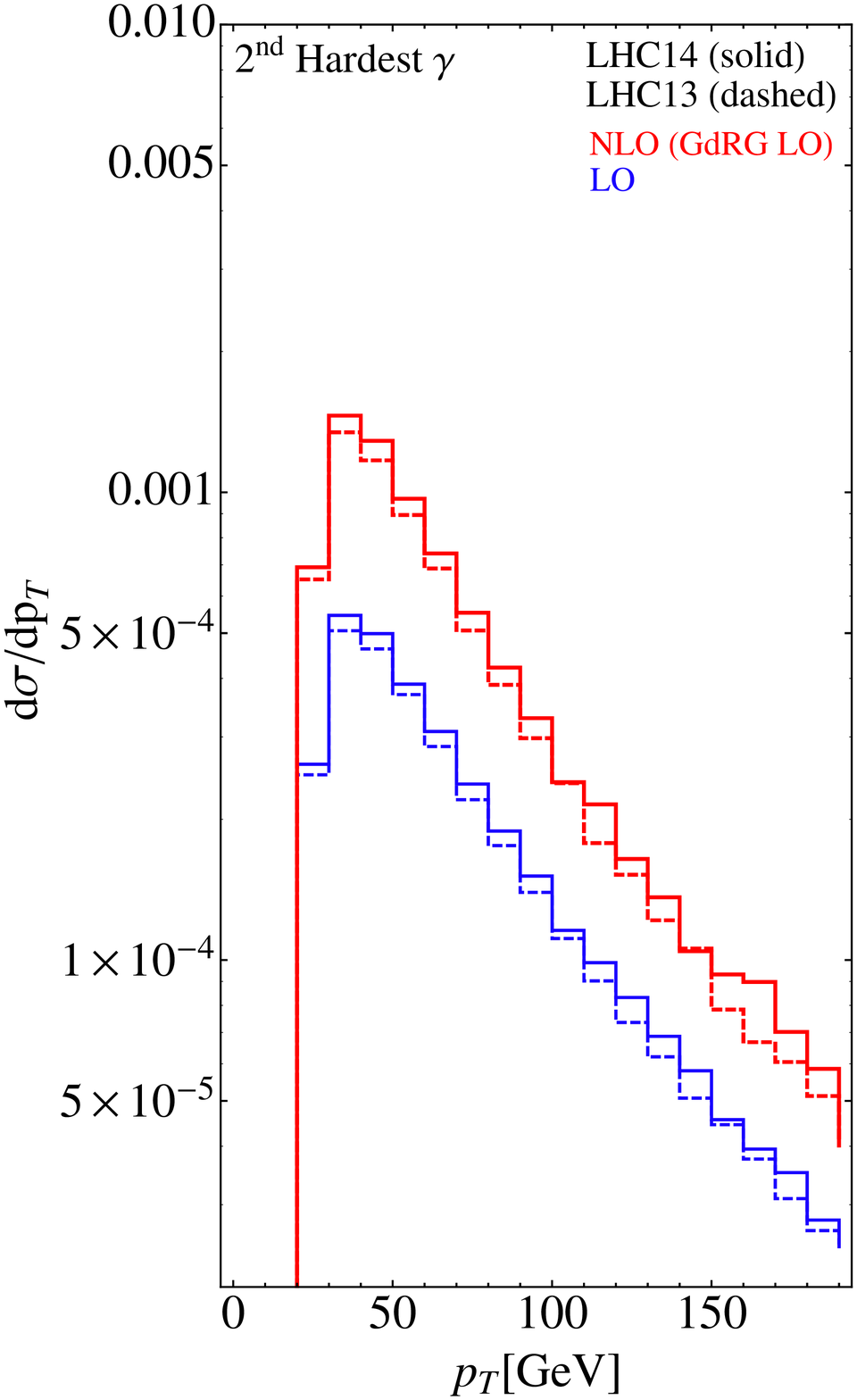}
\includegraphics[width=4cm]{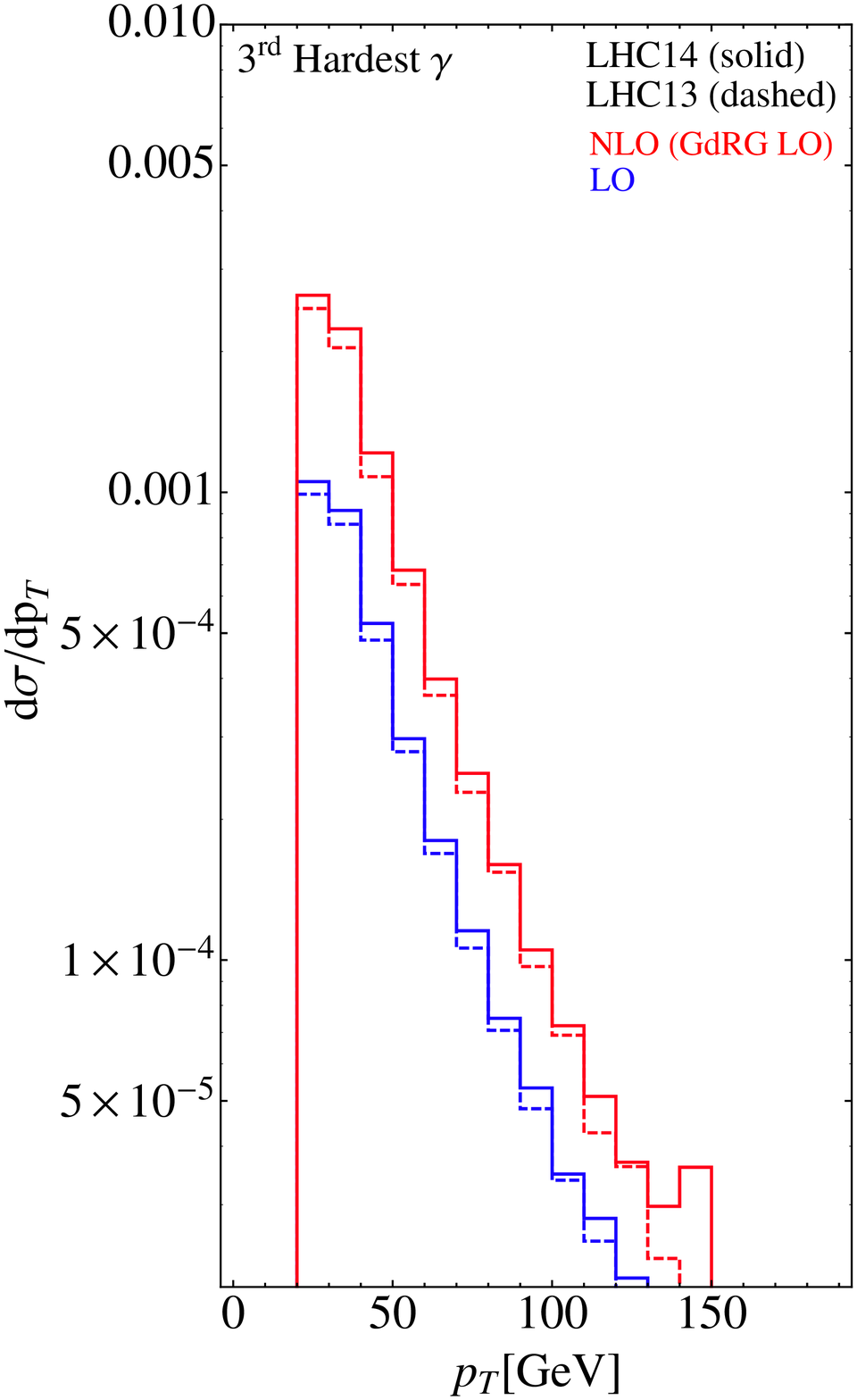}
\includegraphics[width=4cm]{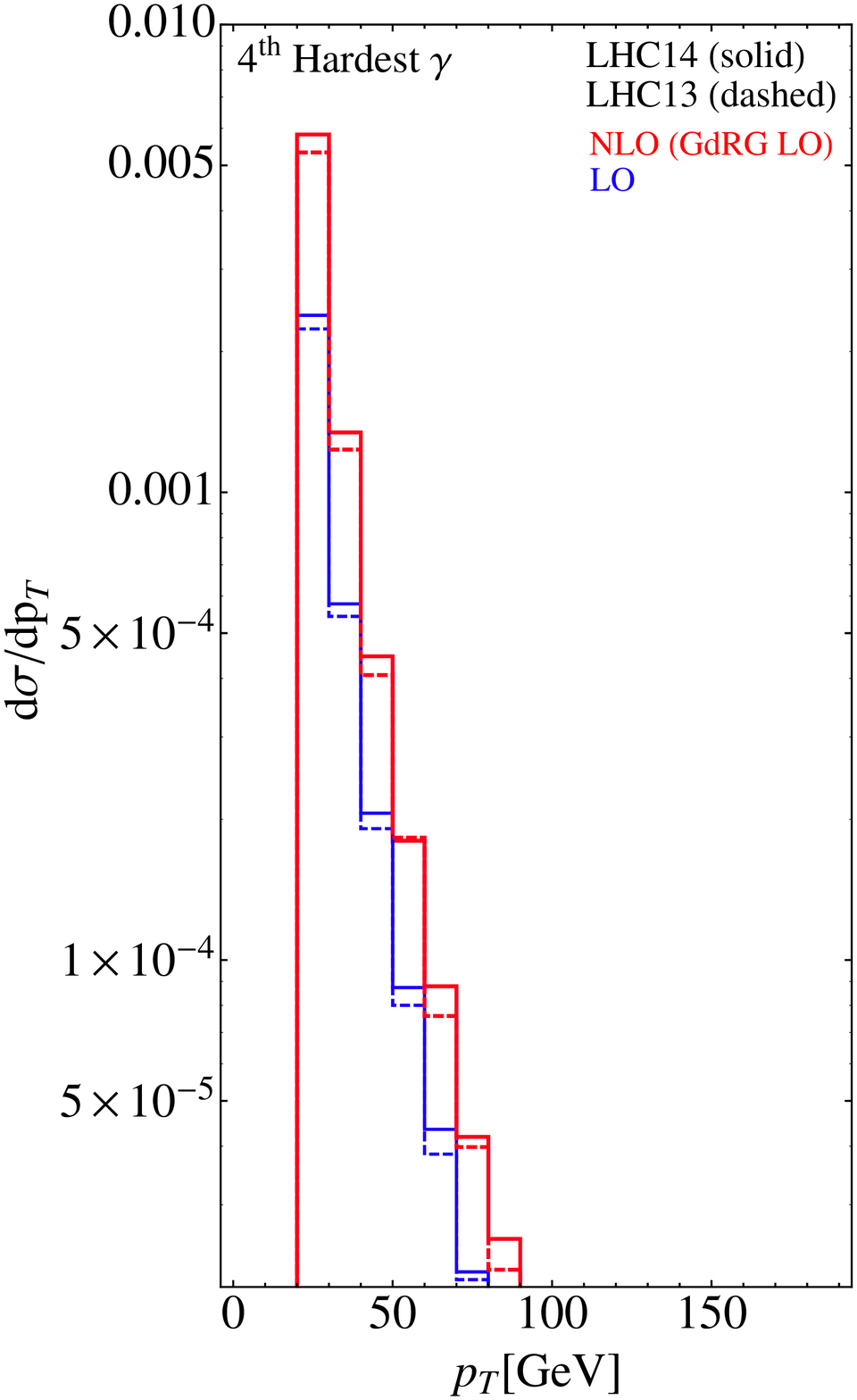} \\
\hspace{0.55cm}
\includegraphics[width=3.4cm]{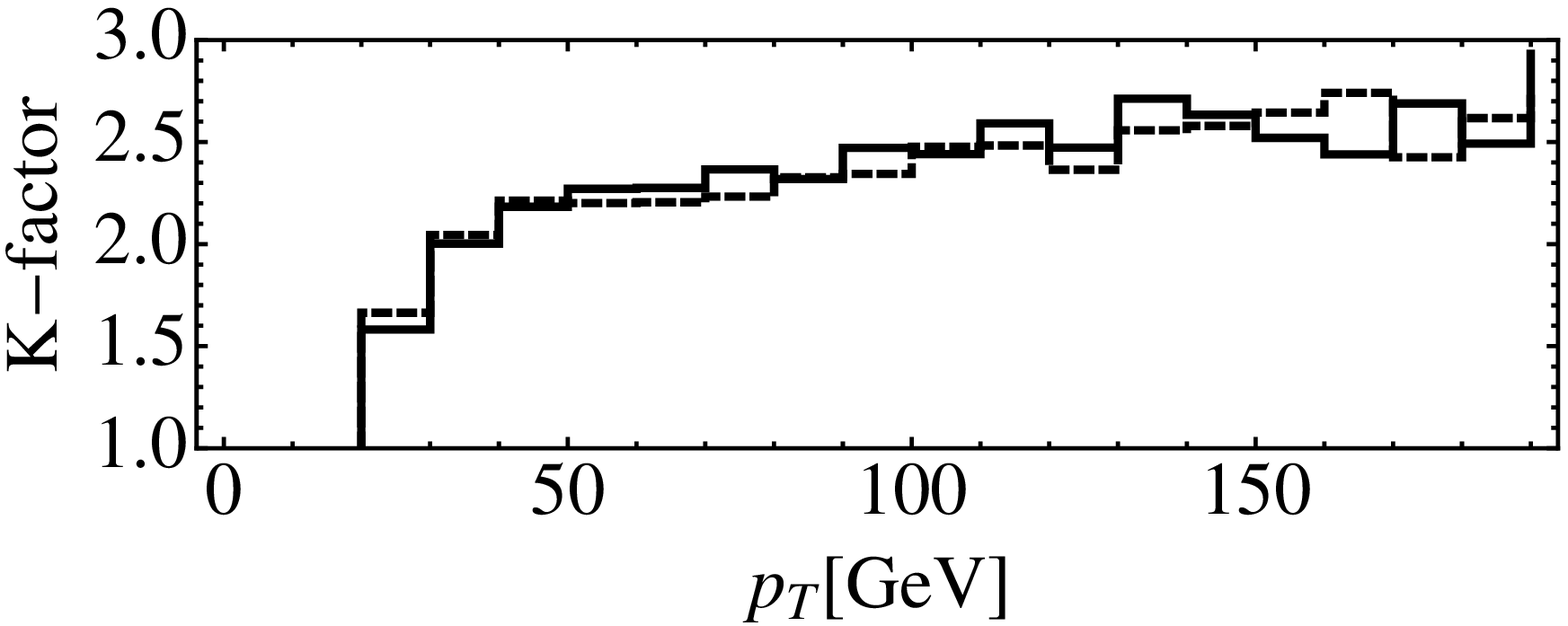}
\hspace{0.55cm}
\includegraphics[width=3.4cm]{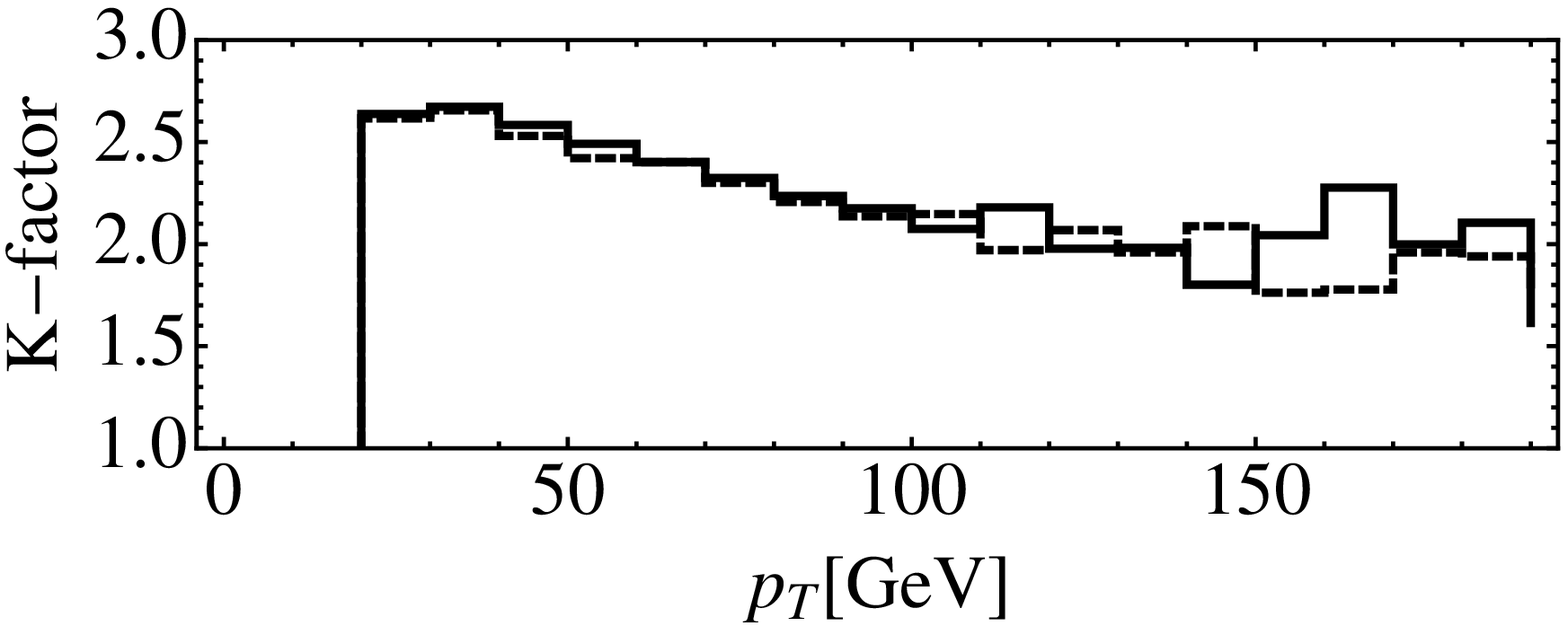}
\hspace{0.58cm}
\includegraphics[width=3.4cm]{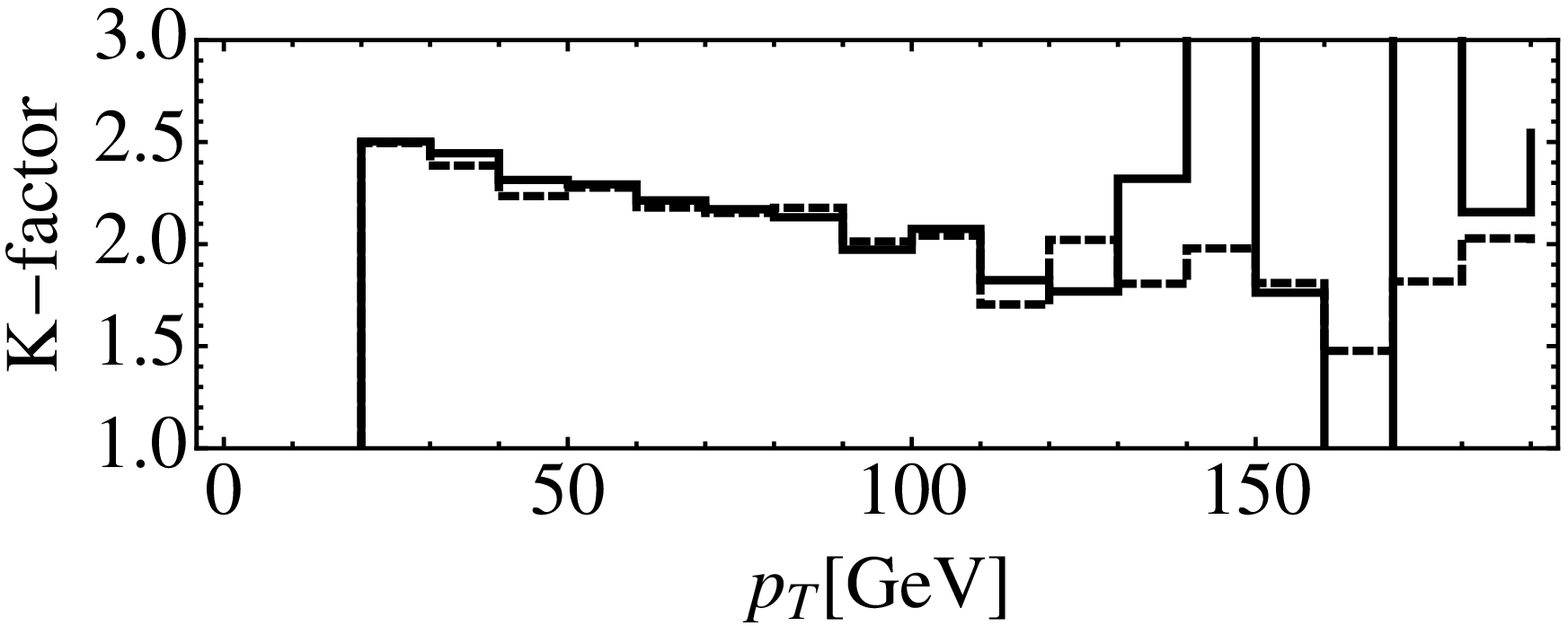}
\hspace{0.55cm}
\includegraphics[width=3.4cm]{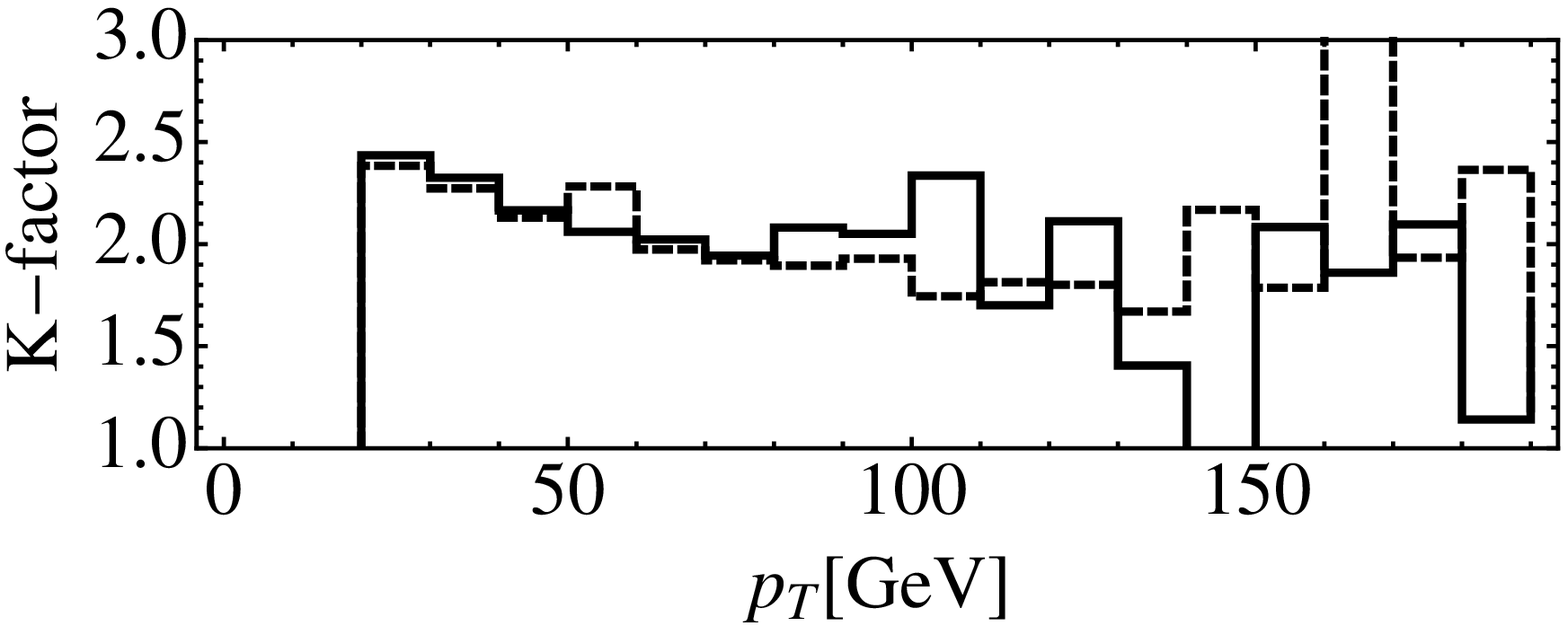} 
\caption{LO and NLO transverse momentum distribution for each of the photons, ranging from hard to soft (left to right). The lower panels
present the differential $K$-factor for each distribution.}
\label{fig:ptdep}
\end{center}
\end{figure}

We present differential distributions for the transverse momentum of the photons in Fig.~\ref{fig:ptdep}. Specifically we investigate the spectrum for 
each of the photons, ranging from the hardest (leftmost) to the softest (rightmost) panel in the figure. Each of the distributions is plotted on the same 
axis to allow for easy comparison. The difference in spread of the $p_T$ values for the four photons is clearly visible. The hardest 
photon spectrum peaks around 50 GeV and has a significant tail. The softest distribution corresponds to a much narrower peak, with essentially all 
the cross section residing in the bins $<  50 $ GeV. This illustrates that even at Run II energies of 13 and 14 TeV, the available phase space for multi-particle production is limited. On the other hand, events with four very hard photons (for instance four photons with $p_T > 100$ GeV) are extremely rare, and any excess of these events could be 
indicative of new physics. Since the differential $K$-factor for the softest photon is relatively flat, it is unlikely that NNLO corrections will induce a significant shape change. 
The hardest photon, however, has a more significant enhancement in the tail at NLO and corresponds to the increase associated with events in which a hard photon recoils against a jet. 
These results may help design the cuts that could be applied in future LHC analyses. One 
possibility is that requiring 4 photons with $p_T > 20$ GeV results in contamination from reducible backgrounds, such as jets faking photons. 
These backgrounds may be suppressed by requiring harder photons. The results of Fig.~\ref{fig:ptdep} illustrate that if one cuts at 40 -- 50 GeV on the hardest (and 
possibly second hardest) photon, the cross section is not dramatically suppressed from the democratic $> 20 $ GeV cut case. 

\subsection{$\gamma\gamma\gamma\gamma$ at future colliders}

An exciting prospect for the future of particle physics is the construction of a new collider with potential center of mass energies of up to 100 TeV~\cite{Avetisyan:2013onh}.
This machine would have fantastic reach, both as a discovery machine and as a probe of existing physics in new regimes. For instance, at these high energies the properties of the EW theory~\cite{Cornwall:1990hh} are such that the amplitudes for the production of $n$ Higgs and $m$ longitudinal vector bosons (at the mass threshold) scale as $\mathcal{A}_{1\rightarrow n+m} \sim n! m!$~\cite{Khoze:2014zha,Khoze:2014kka}. This suggests that for large enough multiplicities the perturbative nature of the EW theory may break down. This breakdown is fascinating to study from both a theoretical and an experimental perspective. The scattering of massless gauge bosons that will remain perturbative may provide a standard candle in this regard, and as such may be an extremely useful process to compare and contrast with the scattering of longitudinal vector bosons. In Table~\ref{tab:high_eng} we present cross sections for $\gamma\gamma\gamma\gamma$ at two choices of $\sqrt{s}$, which may correspond to those of future colliders~\cite{Avetisyan:2013onh}. In addition to the basic cuts we used in the previous section, we also present results for slightly harder cuts, which may be more appropriate at higher energies. At 100 TeV, the cross section is $\mathcal{O}(1)$ fb, which, if a luminosity increase of a couple of orders of magnitude is obtained from the LHC, will result in copious production of these final states, allowing for precision studies of this channel. At higher energies, the $K$-factor has increased from the corresponding value at 14 TeV (2.37). This is due to the dependence of the gluon PDF, which becomes increasingly important at higher energies. Since the LO process is $q\overline{q}$ initiated, it does not experience the same level of enhancement as the NLO cross section. We note that the cross sections presented in the table are likely to be susceptible to significant higher-order corrections at high energies. For example the $gg\rightarrow \gamma\gamma\gamma\gamma$ piece, part of the NNLO calculation, is likely to significantly contribute to the total rate. For example, in the case of $gg\rightarrow ZZ$~\cite{Campbell:2011bn}, at 14 TeV the $gg$ loops represent around 8\% of the LO cross section, whilst at 100 TeV the contribution is 20\%. Therefore it is clear that these pieces will need to be included for phenomenological studies at these energies. Such studies are beyond the scope of this paper, and we leave this for future work.

\begin{table}
\begin{tabular}{|c|c|c|c|c|}
\hline
$\sqrt{s}$ [TeV] & Cuts & LO [fb] & NLO [fb] & $K$-factor \\
\hline
\hline
33  & $p_T^{\gamma} > 20$ GeV & 0.081 & 0.21 & 2.64 \\
      &  $p_T^{\gamma_1} > 50$, $p_T^{\gamma_2} > 40$, $p_T^{\gamma_{3,4}} > 20$ GeV & 0.060 & 0.16 &  2.56\\
\hline
\hline
100  & $p_T^{\gamma} > 20$ GeV & 0.25 & 0.73 & 2.96  \\
      &  $p_T^{\gamma_1} > 50$, $p_T^{\gamma_2} > 40$, $p_T^{\gamma_{3,4}} > 20$ GeV & 0.19 & 0.53  &  2.81 \\
\hline
\end{tabular}
\caption{Cross sections at LO and NLO at $\sqrt{s}$ center of mass energies that may correspond to those of future machines. Cuts are
those described in the previous section, apart from the $p_T$ requirements, which are displayed in the table.
\label{tab:high_eng}}
\end{table}

\section{Conclusions}
\label{sec:conc}

In this paper we have implemented $\gamma\gamma\gamma\gamma$ production at NLO in QCD and studied the
phenomenology of this signal at the LHC. We included the fragmentation functions contribution, allowing for the application 
of experimental style photon isolation requirements. In order to obtain stable and efficient NLO code we used analytic 
unitarity to calculate the coefficients of the one-loop integral functions for each photon helicity configuration. The 
resulting code is stable and can run in double precision mode. It will be released publicly in a forthcoming version of MCFM. The cross sections for this process are small $\mathcal{O}(0.1)$ fb, and as a result require a large data 
set to observe; the full $3000$ fb$^{-1}$ data set should contain hundreds of four photon events. The NNLO contribution $gg\rightarrow 4\gamma$ will likely  
provide a significant increase to the cross section, due to PDF enhancement. We leave this study to future work. 

On the theoretical side, the process $pp\rightarrow \gamma\gamma\gamma\gamma$ serves an excellent testing ground for the applications 
of analytic unitarity to $2\rightarrow 4$ processes. This piece corresponds to the most subleading in color pieces of the $q\overline{q}\rightarrow gggg$ process. 
Over the next few years the NNLO frontier will move to $2\rightarrow 3$ processes, and experience at $2\rightarrow 2$ has highlighted 
the advantages of stable NLO code. We believe the methods of analytic unitarity, applied in this paper to $4\gamma$ production, provide 
an excellent approach to tackling this problem. Our code can be utilized in the calculations of $2\rightarrow 3$ NNLO processes involving photons and jets in the 
final state. We have outlined refinements of some of the existing methods in the literature that will allow for applications involving massive bosons in the future. 

On the experimental side, the study of multiple vector boson production has yielded fantastic insight into the underlying structure of 
the electroweak sector of the SM. By measuring final states with multiple leptons and photons, the non-Abelian structure of the SM can 
be tested and confirmed. By investigating and measuring rare processes, the LHC experiments demonstrate their ability to constrain other
rare processes, which may occur in extensions of the SM. At increasing energies, the differences between the scattering of massive EW bosons and
massless ones becomes extremely interesting to study.

\section*{Acknowledgements} 

We thank Simon Badger, Zvi Bern, John Campbell, Lance Dixon and Harald Ita for useful discussions. TD gratefully acknowledges support from the Danish Council for Independent Research.

\bibliography{Fourgam.bib}
\end{document}